\documentclass[%
 reprint,
superscriptaddress,
 amsmath,amssymb,
 aps,
]{revtex4-2}

\usepackage{graphicx}
\usepackage{dcolumn}
\usepackage{bm}
\usepackage{aas_macros}

\usepackage{hyperref}
\hypersetup{
    colorlinks=true,
    linkcolor=blue,
    filecolor=blue,      
    urlcolor=blue,
    citecolor=blue, 
    pdftitle={GFL paper},
    pdfpagemode=FullScreen,
    }
\usepackage{xcolor}

\begin{document}

\preprint{APS/123-QED}

\title{Efficient large-scale, targeted gravitational-wave\\ probes of supermassive black-hole binaries} 

\author{Maria Charisi}\email{maria.charisi@nanograv.org}
\affiliation{%
 Department of Physics \& Astronomy, Vanderbilt University, 2301 Vanderbilt Place, Nashville, TN 37235, USA
}%
\author{Stephen~R.~Taylor}%
\affiliation{%
 Department of Physics \& Astronomy, Vanderbilt University, 2301 Vanderbilt Place, Nashville, TN 37235, USA
}%

\author{Caitlin A. Witt}%
\affiliation{Center for Interdisciplinary Exploration and Research in Astrophysics (CIERA), Northwestern University, Evanston, IL 60208}
\affiliation{Adler Planetarium, 1300 S. DuSable Lake Shore Dr., Chicago, IL 60605, USA}

\author{Jessie Runnoe}%
\affiliation{%
 Department of Physics \& Astronomy, Vanderbilt University, 2301 Vanderbilt Place, Nashville, TN 37235, USA
}%

\date{\today}

\begin{abstract}

Supermassive black hole binaries are promising sources of low-frequency gravitational waves (GWs) and bright electromagnetic emission. Pulsar timing array searches for resolved binaries are complex and computationally expensive and so far limited to only a few sources. We present an efficient approximation that empowers large-scale targeted multi-messenger searches by neglecting GW signal components from the pulsar term. This Earth-term approximation provides similar constraints on the total mass and GW frequency of the binary, yet is $>100$ times more efficient. 

\end{abstract}

\maketitle

\textit{Introduction.---}Supermassive black hole binaries (SMBHBs) are a natural outcome of galaxy mergers \citep{DeRosa2019}. At the final stages of their evolution, SMBHBs are promising sources of low-frequency gravitational waves (GWs). The most massive binaries (with mass $10^8-10^{10}M_{\odot}$) are targeted by pulsar timing arrays (PTAs), a galactic-scale detector that is sensitive to nanohertz frequencies \citep{Burke-Spolaor2019,2021arXiv210513270T}. PTAs monitor millisecond pulsars, which are stable and provide precisely repeating pulses over decades. Passing GWs from SMBHBs perturb Earth--pulsar separations, inducing deviations in the times of arrival (TOA) of radio pulses. 

PTAs are soon expected to detect a background of GWs from a population of unresolved, inspiraling SMBHBs by measuring a quadrupolar inter-pulsar correlation signature in the timing deviations (i.e., \textit{the Hellings \& Downs curve} \citep{Hellings1983}).  All major PTAs---the North American Nanohertz Observatory for Gravitational waves (NANOGrav) \citep{McLaughlin2013,2019BAAS...51g.195R}, the European Pulsar Timing Array (EPTA) \citep{2013CQGra..30v4009K} and the Parkes Pulsar Timing Array (PPTA) \citep{2008AIPC..983..584M,Hobbs2013}---have detected a red noise process with common spectral properties in many of the monitored pulsars \citep{2020ApJ...905L..34A,2021ApJ...917L..19G,2021MNRAS.508.4970C}. This has also been confirmed by a combination of older datasets from the constituent PTAs of the International Pulsar Timing Array (IPTA) \citep{2022MNRAS.510.4873A}. If this is the first hint of the background, the definitive signature of spatial correlations could be detected in the next few years \citep{2021ApJ...911L..34P}.

Strong GW signals from massive and relatively nearby SMBHBs can be resolved above the GW background and should be detectable within a few years of its detection \citep{Rosado2015,2017NatAs...1..886M,Kelley2018,2020PhRvD.102h4039T,2022ApJ...941..119B}. PTAs will detect SMBHBs thousands of years before coalescence ($\sim10^4$ years for an equal-mass binary with $10^9M_{\odot}$ and period of a few years), which likely show no evolution over the baseline of timing observations ($\sim$~decades). Recent PTA datasets have provided upper limits on the GW strain from individual SMBHBs \citep{2014ApJ...794..141A,Zhu2014,Babak2016,NANOGrav_CWs_2018,2023arXiv230103608A} (with the latter also modeling the common-spectrum process), including stringent mass-ratio limits on tentative SMBHBs in nearby galaxies \citep{2016MNRAS.459.1737S,NANOgrav_Charisi}. 

SMBHBs may reach the final stages of their evolution embedded in gas-rich environments \citep{2005ApJ...620L..79S}. Therefore, in addition to being strong sources of GWs, they likely produce electromagnetic (EM) emission \citep{2022LRR....25....3B}, making them excellent targets for multi-messenger observations \citep{2019BAAS...51c.490K,2022MNRAS.510.5929C}. Combined EM+GW searches are very advantageous as they improve the GW upper limits \citep{2020ApJ...900..102A}, boost the GW detection probability \citep{LiuVigeland2021}, and can significantly improve parameter estimation \citep{LiuVigeland2023,charisi_inprep}. 
Since EM observations provide very precise locations and redshifts, multi-messenger searches are by default targeted towards specific galaxies of interest. This has important implications on the PTA GW data analysis, since the sky location and luminosity distance can be fixed, unlike the typical PTA searches which are all-sky and use a broad, weakly-informative prior for the distance. 

In particular, the timing deviations induced by GWs from a SMBHB include two components: $(1)$ the earth term, which is common in all pulsars (up to directional sensitivity factors) \citep[e.g.,][]{2010arXiv1008.1782C}; and $(2)$ the pulsar terms, which are different in each pulsar, reflecting the GW phase upon passing each one. 
Modeling the pulsar term is crucial for localizing the source \cite{2011MNRAS.414.3251L,2010arXiv1008.1782C}. These searches are complicated and computationally expensive, because, in addition to the standard binary parameters, they also search over multiple parameters for each pulsar (pulsar distance and pulsar-term GW phase) \cite{enterprise}. 
However, for \textit{targeted searches}, in which the candidate source location is determined by EM data, modeling only the Earth term offers a significantly simpler and more efficient alternative.
This will be extremely crucial both for GW follow-ups targeting the flood of EM candidates in the era of the Legacy Survey of Space and Time (LSST) of the Vera Rubin Observatory \citep{2021MNRAS.506.2408X,2021MNRAS.508.2524K,2022ApJ...936...89W}, and for following-up the hundreds or thousands of potential host galaxies within the large localization volume of a GW-triggered detection \citep{2019MNRAS.485..248G}. 
Here, we use realistic binary simulations and compare the parameter estimation and computational efficiency of Bayesian analyses that model the full signal likelihood versus ones that neglect the pulsar terms. The output of our simulations can be found in \url{https://github.com/mariacharisi/Earth-Term-MMA.git}.

\textit{GW Signals and Simulations---}
The timing deviations, $s$, induced by a binary can be written as
  $   s(t, \hat{\Omega}) = F^{+}(\hat{\Omega}) \Delta s_{+}(t) +F^{\times}(\hat{\Omega}) \Delta s_{\times}(t)$, where $\hat{\Omega}$ is the GW unit position vector,
$+, \times$ refer to the GW polarization, $F^{+, \times}$ are the antenna pattern functions that describe the response of an Earth--pulsar system to the GW signal, and  
\begin{equation}
   \Delta s_{+, \times}(t)= s_{+, \times}(t)-s_{+, \times}(t_p)
\end{equation}
is the difference between the Earth term and the pulsar term, which depends on the binary parameters, pulsar distance $L$, pulsar-term GW phase, and $t_p = t - L(1-\hat\Omega\cdot\hat{p})$ where $\hat{p}$ is the unit position vector of the pulsar. The full signal and PTA likelihood function are derived in detail in \citep{NANOGrav_CWs_2018,2023arXiv230103608A}. 

Following the framework developed in \citet{2021ApJ...911L..34P}, we simulate timing data for a PTA with near-future sensitivity. The simulated data have a baseline of 20 years, (roughly the expected timeline for detection of GWs from individually resolved binaries \citep{Rosado2015,2017NatAs...1..886M,Kelley2018,2020PhRvD.102h4039T,2022ApJ...941..119B}), but otherwise resemble the NANOGrav's $12.5$-year dataset \cite{NANOGrav12p5_Dataset} in terms of number of pulsars, observational sampling, TOA uncertainties etc. We also simulate intrinsic red noise for each pulsar according to its measured characteristics in the NANOGrav $12.5$-year dataset.
For the extrapolated data, we keep the observational properties similar to the $12.5$-year dataset (see \citep{2021ApJ...911L..34P} for details). This represents a conservative choice for the future PTA sensitivity, since the number of pulsars will certainly increase.
We inject one SMBHB signal into each of these simulated PTA datasets, randomly drawing the binary parameters from uniform distributions in the following ranges: \textit{sky location}, $\theta: [0, \pi]$, and $\phi: [0, 2\pi]$; \textit{distance}, $\log_{10} (D/\mathrm{Mpc}): [1,3]$; \textit{total binary mass}, $\log_{10} (M_{\rm tot}/M_\odot):[9,10]$; \textit{binary mass ratio}, $\log_{10}q: [-1,0]$; \textit{GW frequency}, $\log_{10} (f/\mathrm{Hz}):[-9,-7.5]$; \textit{orbital inclination angle}, $\cos\iota:[-1,1]$; \textit{initial Earth-term phase}, $\Phi_0: [0,2\pi]$; and \textit{ GW polarization angle}, $\psi: [0,\pi]$.
The simulated timing deviations include the pulsar terms with pulsar distances 
from \citet{pulsar_dist},
and allow for frequency evolution between the Earth and pulsar terms but not within the timing baseline of the data. 
We simulate $1,500$ binary signals, and compute the signal-to-noise ratio (S/N) for each. We exclude realizations with $\mathrm{S/N}<5$  and $\mathrm{S/N}>100$, since our goal is to test our analysis approximation in moderate signal regimes where any biases would show, yet not so strong as to be unrealistic for PTAs. This leaves $824$ injected-binary datasets remaining in the sample.

For each simulated dataset, we perform  Bayesian analyses using the \texttt{enterprise} PTA software \cite{enterprise}, and \texttt{PTMCMCSampler} \citep{ptmcmc} to sample the posterior distribution of the parameter space, employing uniform priors with ranges like above. However, for the GW frequency and total mass we use priors that are standard for PTA searches, i.e.  $[-7,-9]$ and $[6,10]$, respectively. We exclude high frequency binaries from our simulations, because they are rare and unlikely to be detected due to limited sensitivity at higher frequencies \citep{Rosado2015,2017NatAs...1..886M,Kelley2018,2020PhRvD.102h4039T,2022ApJ...941..119B}.
Similarly, we inject only high-mass binaries, since PTAs are currently sensitive only to the most massive systems \citep{NANOgrav_Charisi} and to ensure high $\mathrm{S/N}$, but our conclusions should also hold for lower-mass binaries. We perform two sets of analyses on each dataset:\\
$(i)$ The full signal model, which includes the pulsar terms (\textit{PTerm} hereafter). This setup includes $6$ binary parameters 
plus $90$ pulsar parameters (pulsar distance, pulsar-term GW phase for each).
To effectively sample the high-dimensional parameter space, we begin the chains at the correct binary parameters (to emulate the outcome of an initial search stage), and include dedicated Markov chain Monte Carlo (MCMC) proposal distributions designed for PTA SMBHB analyses \citep{11yrCW}.\\
$(ii)$ A simplified model, which includes the Earth term but excludes all pulsar terms (\textit{ETerm} hereafter). This setup needs only the $6$ parameters describing the binary waveform. Here we start the samplers at random initial positions within our priors, and do not use sophisticated proposal distributions.\\
Since we consider targeted GW searches (e.g., as in \citep{2020ApJ...900..102A,LiuVigeland2021,LiuVigeland2023}) in both analyses the sky position and distance of the source are fixed to the injected value. Likewise, we fix the per-pulsar red noise characteristics to injected values since this will not alter the generality of our results.

\begin{figure}
\includegraphics[width=\columnwidth]{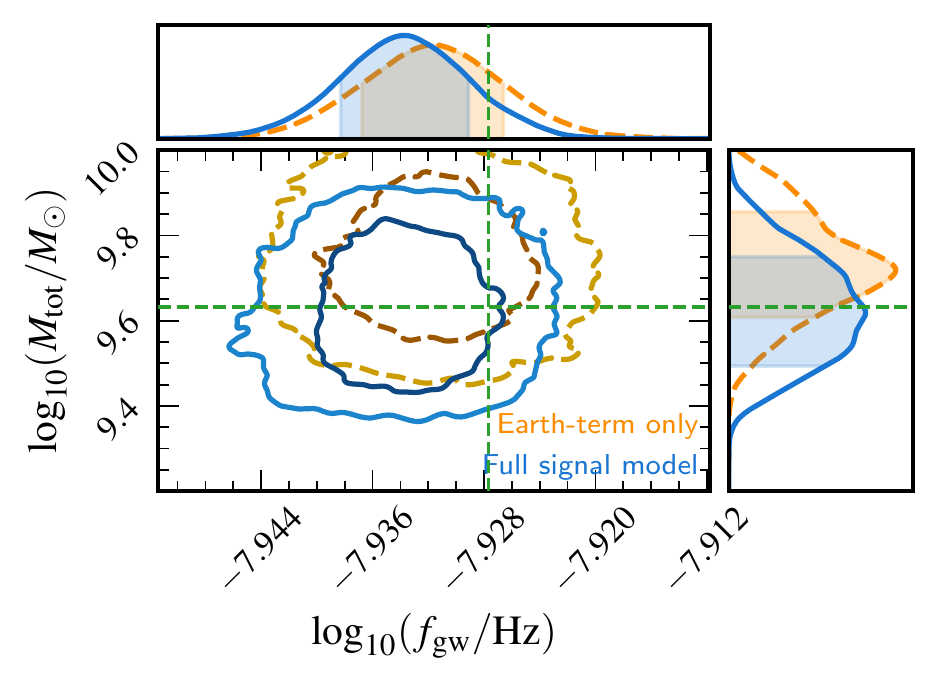}
\caption{\label{fig:cornerplot}Posterior distributions of the total mass and GW frequency. The \textit{ETerm} and \textit{PTerm} analyses are shown with orange dashed and blue solid lines, respectively, with green dashed lines showing the injected values. The shaded regions in the 1D posteriors show the 68\% credible intervals.}
\end{figure}

\begin{figure*}
\includegraphics[width=\textwidth]{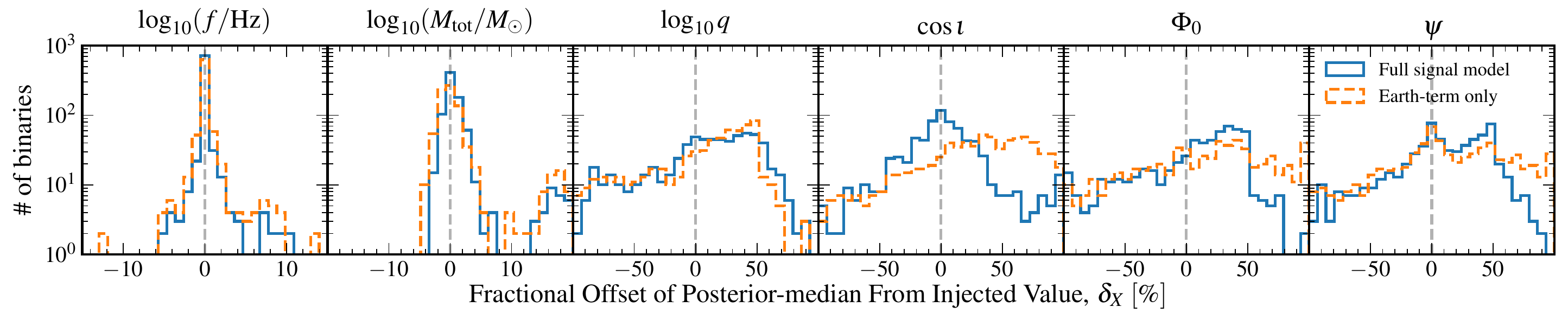}
\includegraphics[width=\textwidth]{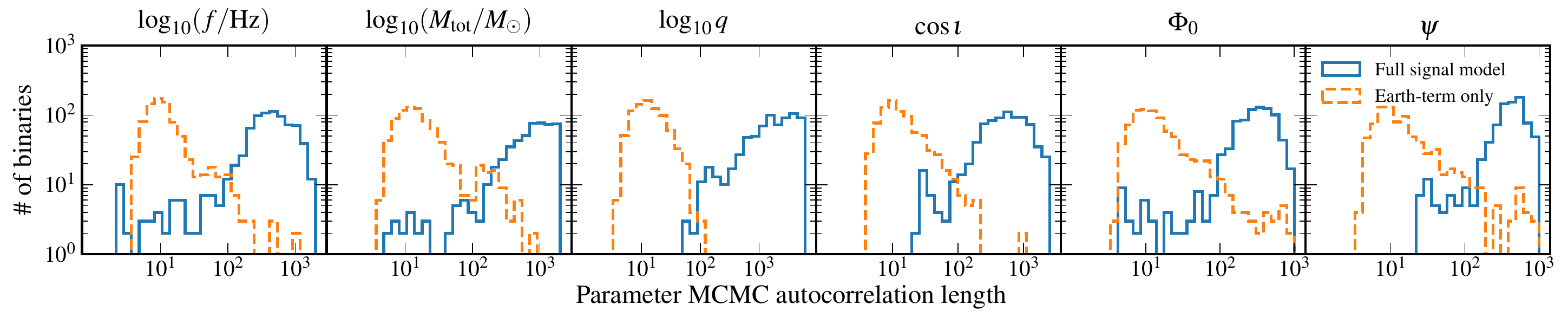}
\caption{\label{fig:Fractional_Error} Comparison of the \textit{ETerm} (orange dashed lines) and \textit{PTerm} (blue solid lines) analyses in terms of parameter estimation (top) and computational efficiency (bottom). The top panel shows the distribution of percent error, $\delta_X [\%]$ of the posterior median with respect to the injected value with vertical grey lines delineating 0\%. The bottom panel shows the distribution of autocorrelation length $L_x$ of the MCMC chains.}
\end{figure*}

\textit{Results.---} 
In \autoref{fig:cornerplot}, we show an example of the marginalized posterior probability distributions for the total mass and GW frequency for a binary  with $\mathrm{S/N}\sim9$. We see that both analyses successfully recover the injected values within $68\%$ credibility, and return similar constraints on these parameters. Next, we systematically examine how well the injected values, $X_{\rm in}$, are recovered across all our simulated datasets, by calculating the percentage errors of the posterior median 
$\delta_X=(X_{\rm in}-X_{{\rm post.},50})/X_{\rm in}\times100\%$, where $X$ is any of the six binary parameters. In the top panel of \autoref{fig:Fractional_Error}, we show the distribution of the percentage error for each binary parameter both with the \textit{ETerm} (orange dashed lines) and \textit{PTerm} (blue solid lines) analyses. 
In ~\autoref{Tab:FE}, we show the $16^{\rm th}$, $50^{\rm th}$, and $84^{\rm th}$ percentiles of each error distribution. We also report the percentage of realizations, $P_{10\%}$, for which the posterior median is within 10\% of the injected value, i.e., $|\delta_X| < 10\%$.

Both methods successfully recover the GW frequency and binary total mass, with narrow $\delta_X$ distributions peaked around $0\%$ and high $P_{10\%}$ percentages. The \textit{PTerm} analysis slightly outperforms the \textit{ETerm} analysis in the case of the binary mass ratio, but even the \textit{PTerm} analysis does not provide tight constraints, as evidenced by the wide distributions of $\delta_X$ and the relatively low $P_{10\%}$ percentages. At leading order, the evolution of the GW waveform depends on the binary \textit{chirp mass} $M_c=q^{3/5}/(1+q)^{6/5}M_{\rm tot}$. Therefore, the \textit{PTerm} analysis performs better because it can more effectively constrain the evolution of the chirp mass (which depends on the mass ratio) via the pulsar terms, which reflect the GW perturbations at the location of the pulsars thousands of years ago. The binary inclination is also better constrained by the \textit{PTerm} analysis. 
We note that for multi-messenger observations, the inclination and mass ratio may be independently constrained by the EM signal \citep{2022MNRAS.510.5929C}, and thus the simpler \textit{ETerm} analysis may be sufficient even for these parameters \citep{charisi_inprep}. Finally, the initial Earth-term phase and GW polarization angle are poorly constrained in both analyses (see the \textit{Discussion} below).

\begin{table}
\caption{\label{Tab:FE} Percentiles ($16^\mathrm{th}$, $50^\mathrm{th}$ and $84^\mathrm{th}$) of the percent error distributions, $\delta_X$, and fraction of realizations, $P_{10\%}$, for which the median is within 10\% of the injected value for each binary parameter and for each analysis.}

\begin{ruledtabular}

\begin{tabular}{lllll}
&\textit{PTerm}&\textit{ETerm}&\textit{PTerm}&\textit{ETerm}\\
parameter&$\delta_{X}$ [\%]& $\delta_{X}$ [\%]& $P_{10\%}$ [\%]& $P_{10\%}$ [\%]\\
\hline
$\log_{10} f$ & $0.0_{-0.1}^{0.1}$ &$0.0_{-0.2}^{0.4}$&99.8&99.4\\
$\log_{10} M_{\rm tot}$ &  $0.3_{-0.6}^{1.8}$&$-0.2_{-1.6}^{1.9}$&95.5&93.0\\
$\log_{10} q$ &$4.4_{-137.9}^{45.9}$&$8.9_{-173.2}^{43.6}$&15.3&10.2 \\
$\cos i$& $0.7_{-34.1}^{32.0}$&$36.9_{-27.1}^{81.5}$&33.1&9.7\\
$\Phi_0$&$1.9_{-243.5}^{40.4}$&$-4.9_{-251.7}^{58.4}$&10.9& 9.8\\
$\psi$& $0.1_{-174.3}^{44.2}$&$-0.4_{-168.9}^{57.0}$&18.9&18.2\\
\end{tabular}
\end{ruledtabular}

\end{table}

The above metric for parameter-estimation fidelity relies on the posterior median, i.e., a point estimate, but the shape of the posterior distribution contains more information. We further compare the two analyses by estimating the ratio $R_{\Delta X}=\Delta X_{68} (\mathit{ETerm})/\Delta X_{68} (\mathit{PTerm})$ of the $68\%$ credible intervals $\Delta X_{68}=X_{84}-X_{16}$ for each parameter $X$. In \autoref{Tab:Rx}, we show the median of the $R_{\Delta X}$ distribution for each parameter. We see that the \textit{ETerm} analysis provides $4\%$ wider constraints on the GW frequency, $25\%$ wider for the binary total mass, and $18\%$ wider for the mass ratio. On the other hand, the \textit{ETerm} analysis provides $15\%$ narrower credible intervals for the orbital inclination, and significantly narrower $\Delta X_{68}$ for the initial Earth-term phase and the GW polarization angle (but also see the \textit{Discussion}).

Next, we calculate the Kullback–Leibler divergence $D_{KL}$, which measures the difference in information content between two distributions, a parameter's prior and posterior in our case (e.g., see \citep{LiuVigeland2021} for a detailed description of this metric). 
A high value of $D_{KL}$ shows more deviation of the posterior from the prior (and thus higher gain of information from the data), while a value of zero signifies identical distributions (i.e., the data did not update the prior information). 
In \autoref{Tab:Rx}, we report the median of the distribution of $D_{KL}$ ratios between the two analyses, $R_{KL_X}=D_{KL_X} (\mathit{ETerm})/D_{KL_X} (\mathit{PTerm})$. Based on this ratio, we see that both analyses perform similarly for the GW frequency and binary total mass, with the \textit{PTerm} returning on average $1\%$ higher $D_{KL}$ values for the GW frequency, and $8\%$ higher for the binary total mass, which are consistent with our findings above. On the other hand, $D_{KL}$ is 50 times higher in the \textit{PTerm} analysis for the mass ratio. This is driven by a large fraction of realizations ($\sim80\%$) for which the  posterior is almost identical to the prior (with $D_{KL}<1$) in the \textit{ETerm} analysis. For the remaining parameters, $D_{KL}$ is on average higher for the \textit{ETerm} analysis (again see the \textit{Discussion} for more on this).

As a final test of the fidelity of the \textit{ETerm} analysis for parameter estimation, we quantify potential biases using a $p$--$p$ test. We assess the fraction of our simulated binary datasets for which the injected value falls within a given $p\%$ credible interval, $P_p$, where unbiased coverage would return $P_p=p$. We note that since our ranges of injected binary parameters differ slightly from our analysis priors for total mass and GW frequency, we do not expect perfectly unbiased coverage even for the \textit{PTerm} analysis. 
In \autoref{Tab:Rx},
we show the $P_{68}$ and $P_{95}$ values for both analyses. 
The \textit{ETerm} analysis performs better than the \textit{PTerm} analysis for the total mass and mass ratio, with $P_{68}$ and $P_{95}$ within 10\% of the expected values, and slightly worse for the GW frequency ($P_{95}$ is within 15\% from the expected). However, in the three angular parameters we see significant deviations from the expected values in the \textit{ETerm} analysis, especially for $P_{95}$, which points to significant bias in the parameter estimation. However, as we explain in the \textit{Discussion}, this is not significant for targeted searches, since these parameters may be constrained by the EM signal \citep{2022MNRAS.510.5929C} and are not important for finding potential host galaxies.

\begin{table}
\caption{\label{Tab:Rx} Median values for the distributions over simulations of the 68\% credible interval ratio, $R_{\Delta X}$, the KL Divergence ratio, $R_{KL}$, and the auto-correlation length ratio, $R_L$. The last two columns show the $P_{68}$ and $P_{95}$ values (\textit{ETerm}$|$\textit{PTerm}), i.e. the value for the \textit{ETerm} analysis shown on the left and the respective value for \textit{PTerm} on the right.}
\begin{ruledtabular}
\begin{tabular}{lllllll}
parameter&$R_{\Delta X}$& $R_{KL}$& $R_L$ &$P_{68}$ &$P_{95}$ \\
\hline
$\log_{10} f$ & $1.04$&$0.99$& 43.0& 62$|$74& 81$|$95\\
$\log_{10} M_{\rm tot}$ & $1.25$ & $0.92$& 94.5& 68$|$52& 89$|$80 \\
$\log_{10} q$ & $1.18$&$0.02$& 150.3&70$|$66 &97$|$90 \\
$\cos i$& $0.85$&$1.22$&39.6&60$|$68 &76$|$95 \\
$\Phi_0$&$0.22$&$3.04$&24.5&59$|$68& 71$|$90 \\
$\psi$& $0.10$&$2.72$&33.7&60$|$67&72$|$94\\
\end{tabular}
\end{ruledtabular}
\end{table}

Finally, we compare both methods in terms of computational efficiency. For this, we calculate the auto-correlation length $L_{X}$ for each parameter $X$, which quantifies how often independent samples are drawn in the MCMC chains. In the bottom panel of \autoref{fig:Fractional_Error}, we show the distributions of $L_{X}$ for the \textit{ETerm} and \textit{PTerm} analyses. Note that we thinned the chains by a factor of 10 and $L_{X}$ is calculated in the final chains. In \autoref{Tab:Rx}, we report the median of the ratio of autocorrelation lengths between the two analyses $R_L=L_X(PTerm)/L_X(ETerm)$. We see that, depending on the parameter, the \textit{PTerm} analysis requires between 25 to 150 more steps to draw independent samples. Finally, in order to ensure overall convergence in the analysis, every parameter needs to have enough independent samples, and thus the total length of the MCMC chain is determined by the chain with the longest $L_X$. In order to compare the efficiency of the two analyses, we calculate the maximum $L_X$ among the 6 binary parameter chains, $L^{\rm max}$, and then calculate the ratio of the maxima for the two analyses $R_{L^{\rm max}}=L^{\rm max}(PTerm)/L^{\rm max}(ETerm)$. This provides an estimate of how much longer the \textit{PTerm} analysis must be run in order to collect the same number of independent samples. The 16$^{\rm th}$,  50$^{\rm th}$, and 84$^{\rm th}$ percentiles of the $R_{L^{\rm max}}$ distribution are $119.1_{14.0}^{330.5}$. Therefore, we conclude that the \textit{ETerm} analysis requires $\sim$120 times fewer steps and thus is $\sim$120 times more efficient. We also compare the memory and CPU requirements of the above runs. On average, the \textit{PTerm} analysis is $\sim$8 times more memory intensive, takes $\sim$7 times longer to complete and requires $>$5 times more disk space for the output chains. Beyond these quantitative comparisons, the \textit{ETerm} is overall significantly simpler and easier to set up. For instance, the \textit{PTerm} analysis requires fine tuning, e.g., including MCMC proposal distributions and starting the chains at the correct binary parameters. Even though the latter is to emulate an initial search, it is too optimistic that such a search would provide the exact parameters. Realistically, the \textit{PTerm} analysis would require further fine tuning (or more steps), which strengthens our conclusions about the comparison of the two analyses.

\textit{Discussion.---}
Searches for individually-resolved supermassive black-hole binaries are among the most complicated and computationally expensive PTA analyses. So far, they have only been possible for a small number of targets \citep{2020ApJ...900..102A, NANOgrav_Charisi}. The problem will be exacerbated in future PTA datasets, since the higher number of pulsars will inevitably increase the dimensionality of the parameter space, and in turn the computational demands. 

Currently, it is intractable to perform a systematic campaign of targeted searches for all SMBHB candidates identified in time-domain surveys ($\sim 250$ systems) \citep{2022MNRAS.510.5929C}, and soon the vast photometric dataset of the Rubin Observatory will potentially uncover thousands of SMBHB candidates \citep{2021MNRAS.506.2408X,2021MNRAS.508.2524K,2022ApJ...936...89W}. Similarly, the first PTA detection of an individually-resolved binary---with its poor localization of potentially hundreds of square degrees---will allow for many potential host galaxies in its error volume \citep{2019MNRAS.485..248G}. Targeted multi-messenger follow-ups of EM identified candidates or promising host-galaxy candidates require efficient and reliable alternatives to the traditional pipeline. This led to the recent development of \texttt{QuickCW} \citep{2022PhRvD.105l2003B}, which delivers an accelerated Bayesian analysis by restructuring the exploration of the likelihood function.  

Here we present a simpler possibility, which will enable systematic large-scale multi-messenger studies of SMBHBs. 
Our comprehensive comparison of targeted GW searches demonstrates that the simplified and significantly more efficient \textit{ETerm} analysis can provide comparable constraints with the more complex and computationally demanding \textit{PTerm} analysis. Both searches return similar constraints on the total mass and GW frequency of the binary, with the posterior median being within $10\%$ of the injected value for the vast majority of realizations (over $90\%$ for both analyses and both parameters). Despite evidence for slight deviations from a completely unbiased estimation based on the $p$--$p$ test, the level of bias is adequate and manageable for these parameters, since supermassive black-hole mass measurements available in galaxy catalogs typically have even higher uncertainty \citep{NANOgrav_Charisi}. Moreover, if the targeted binary candidate is identified as a quasar with periodic variability \citep{Graham2015,Charisi2016,Liu2019,Chen2020}, the uncertainty on the period can also be significant because of the intrinsic stochastic variability, especially when only a few cycles of periodicity are observed. The remaining parameters are not particularly well constrained in either analysis, but the \textit{PTerm} analysis performs slightly better for binary mass ratio and orbital inclination. This is not a major limitation because these parameters may be independently constrained from the EM data for the case of EM candidates \citep{2022MNRAS.510.5929C}, while they are less important for host galaxy identification.

As mentioned, the initial Earth-term phase and GW polarization angle are not well constrained in either analysis. These two parameters are degenerate, because signals with ($\Phi_0$, $\psi$) and ($\Phi_0+\pi$, $\psi+\pi/2$)
produce identical TOA deviations, which may result in bimodal $2$-D posterior distributions in these parameters. Such posteriors are observed in the \textit{PTerm} analysis, but less often in the \textit{ETerm} analysis. The tuned MCMC proposal distributions employed in the former likely force the sampler to more aggressively explore these parameters, whereas in the minimally-tuned \textit{ETerm} analysis the sampler may get stuck in one of the modes. The bimodality of posteriors in the \textit{PTerm} analysis can explain the higher uncertainties $\Delta\Phi_0$ and $\Delta\psi$ for \textit{PTerm}, as well as the lower $D_{\rm KL}$ values; the \textit{ETerm} posteriors are more peaked and thus deviate more from the uniform prior distributions.

Finally, we note that while we performed realistic simulations of near-future PTA sensitivity based on the NANOGrav $12.5$-year dataset, we only injected GW signals from single resolvable binaries. Theoretical models predict that such binaries will likely be detected after the GW background \citep{Rosado2015,2017NatAs...1..886M,Kelley2018,2020PhRvD.102h4039T,2022ApJ...941..119B}, hints of which may be also present in existing datasets  \citep{2020ApJ...905L..34A,2021ApJ...917L..19G,2021MNRAS.508.4970C}. In future simulations, we will also explore whether our results hold in the presence of a stochastic GW background, and in the presence of signals from other resolved binaries.

\textit{Summary.---} With realistic simulations that emulate near-future PTA sensitivity, we compared the performance of targeted GW searches using a full signal analysis, \textit{PTerm}, and a simpler and faster \textit{ETerm} approximation, which neglects the pulsar terms. We found that the \textit{ETerm} analysis provides similar constraints on the binary total mass and GW frequency, and is $>$100 times more efficient. This analysis acceleration empowers the rigorous targeted examination of large samples of candidate SMBHB systems, many of which have already been found, and many more of which are promised by the advent of new time-domain surveys like LSST. This method can also be applied to select promising host galaxies in the large error volume of the first individually resolved binary detected by PTAs.

\begin{acknowledgments}
We thank Nihan Pol, Polina Petrov, William Lamb and our colleagues in NANOGrav and the International Pulsar Timing Array for fruitful discussions and feedback during the development of this technique. MC, SRT, JR acknowledge support from NSF AST-2007993. SRT also acknowledge support from the NANOGrav NSF Physics Frontier Center \#2020265, and an NSF CAREER \#2146016. CAW acknowledges support from CIERA, the Adler Planetarium, and the Brinson Foundation through a CIERA-Adler postdoctoral fellowship. This work was conducted in part using the resources of the Advanced Computing Center for Research and Education (ACCRE) at Vanderbilt University, Nashville, TN. This work was performed in part at Aspen Center for Physics, which is supported by National Science Foundation grant PHY-2210452.
\end{acknowledgments}

\appendix

\bibliography{apssamp}

\end{document}